\documentclass[3p,times,procedia]{elsarticle}
\usepackage{nupha_ecrc}

\usepackage[figuresright]{rotating}
\usepackage{bm,amsmath,amssymb}
\usepackage[mathscr]{eucal}
\usepackage{color}

\long\def\comment#1{ }
\newcommand{\eqn}[1]{Eq.~\eqref{#1}}
\newcommand{\beq}{\begin{equation}}
\newcommand{\eeq}{\end{equation}}

\newcommand{\dif}{{\rm d}}
\newcommand{\rmd}{{\rm d}}

\newcommand{\rmJ}{{\rm J}}
\newcommand{\del}{\partial}

\newcommand{\bk}{\bm{k}}

\newcommand{\bp}{\bm{p}}
\newcommand{\bx}{\bm{x}}
\newcommand{\by}{\bm{y}}
\newcommand{\bu}{\bm{u}}

\newcommand{\bz}{\bm{z}}

\newcommand{\abar}{\bar{\alpha}}

\newcommand{\sdla}{{\rm \scriptscriptstyle DLA}}

\newcommand{\Nc}{N_{\rm c}}

\newcommand{\Nf}{N_{\rm f}}

\newcommand{\minus}{\!-\!}

\volume{00}

\firstpage{1}

\journalname{Nuclear Physics A}

\runauth{E.~Iancu et al.}

\jid{nupha}

\jnltitlelogo{Nuclear Physics A}

\begin{document}

\begin{frontmatter}



\dochead{}

\title{The collinearly-improved Balitsky-Kovchegov equation}


\author[sac]{E.~Iancu}

\author[sac]{J.D.~Madrigal}

\author[col]{A.H.~Mueller}

\author[sac]{G.~Soyez}

\author[ect]{D.N.~Triantafyllopoulos}

\address[sac]{Institut de Physique Th\'{e}orique, CEA Saclay, CNRS UMR 3681, F-91191 Gif-sur-Yvette, France}

\address[col]{Department of Physics, Columbia University, New York, NY 10027, USA}

\address[ect]{European Centre for Theoretical Studies in Nuclear Physics and Related Areas (ECT*)\\and Fondazione Bruno Kessler, Strada delle Tabarelle 286, I-38123 Villazzano (TN), Italy}

\begin{abstract}
The high-energy evolution in perturbative QCD suffers from a severe lack-of-convergence problem,
due to higher order corrections enhanced by double and single transverse logarithms. We resum double logarithms to all orders within the non-linear Balitsky-Kovchegov equation, 
by taking into account successive soft gluon emissions strongly ordered in lifetime. We further resum single logarithms generated by the first non-singular part of the splitting functions and by the one-loop running of the coupling. The resummed BK equation admits stable solutions, which are used to successfully fit the HERA data at small $x$ for physically acceptable initial conditions and reasonable values of the fit parameters. 
\end{abstract}




\end{frontmatter}



\section{The BK equation: from LO to NLO}
\label{sec:bk}

\begin{figure}[t] \centerline{
\includegraphics[width=0.25\textwidth,angle=0]{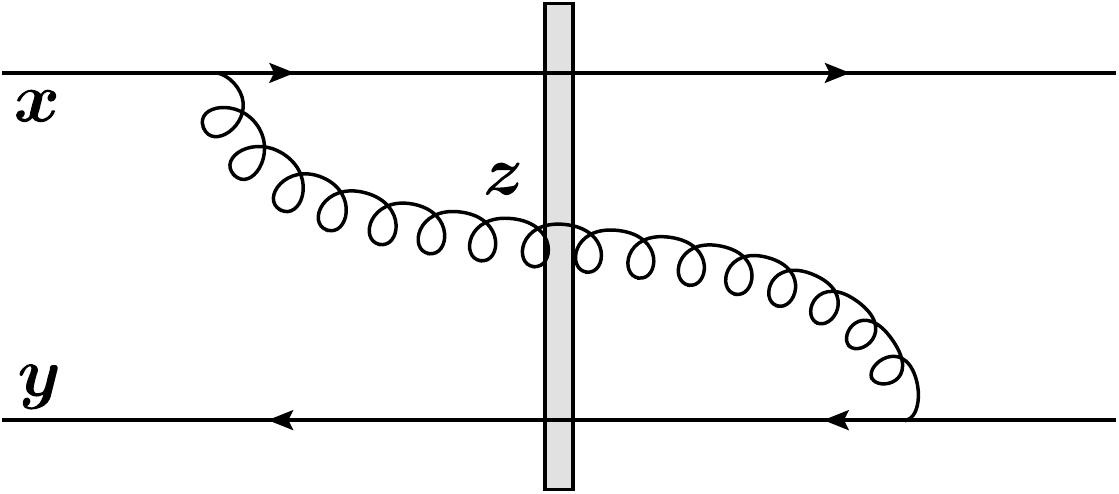}\qquad
\includegraphics[width=0.25\textwidth,angle=0]{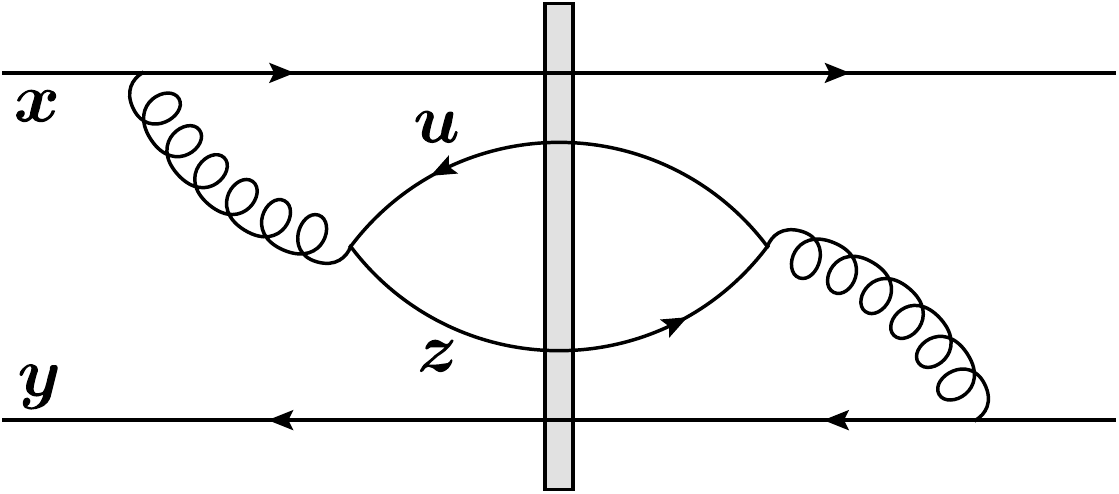}\qquad
\includegraphics[width=0.25\textwidth,angle=0]{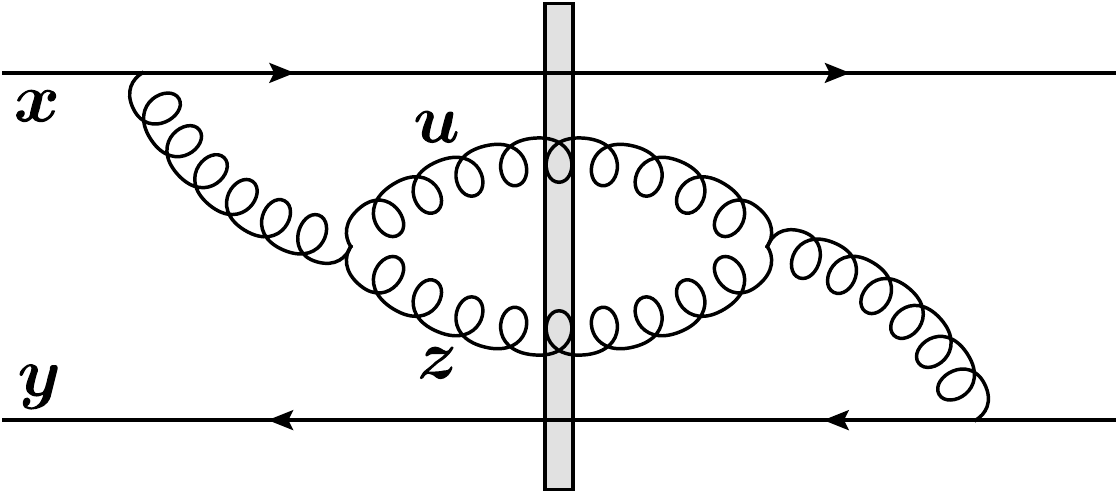}}
\vspace{0.25cm}
\centerline{
\includegraphics[width=0.25\textwidth,angle=0]{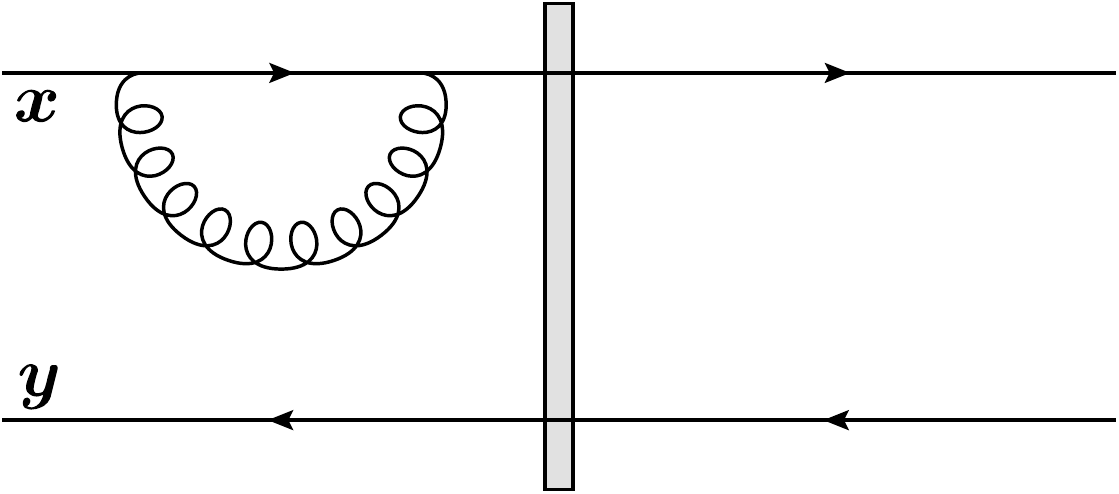}\qquad
\includegraphics[width=0.25\textwidth,angle=0]{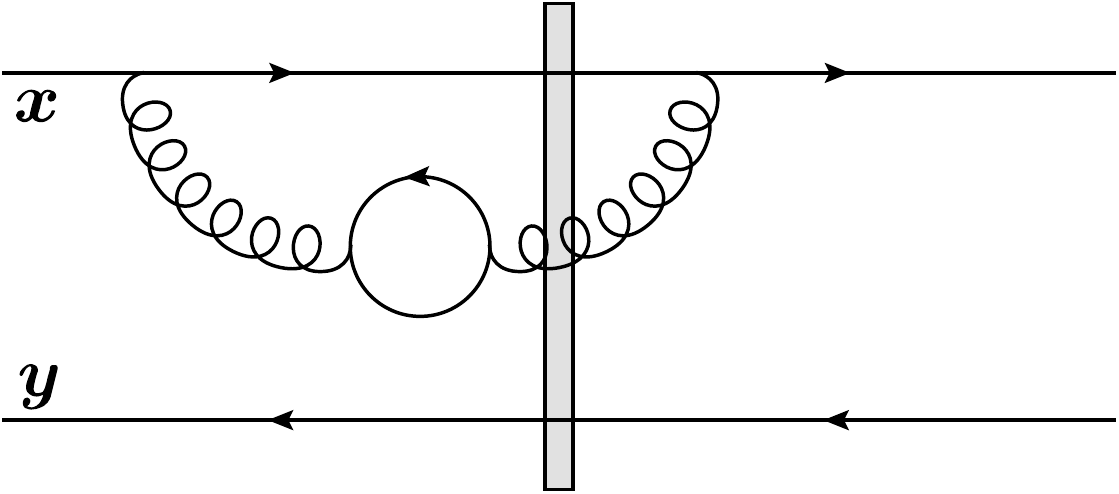}\qquad
\includegraphics[width=0.25\textwidth,angle=0]{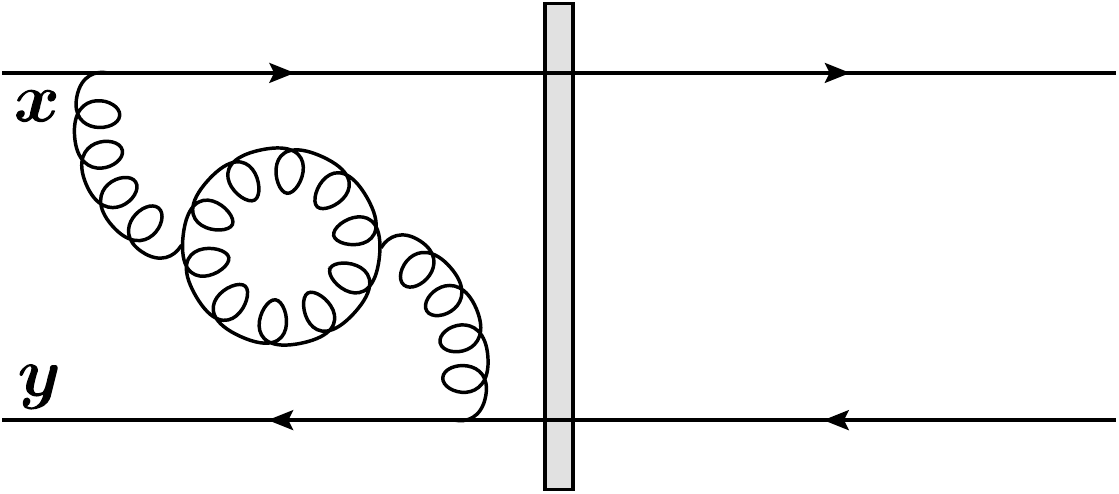}}
\caption{\label{fig:diag} Typical diagrams contributing to the BK equation. The thick vertical line stands for the hadronic target, depicted as a shockwave. Left: LO terms. Middle: NLO terms 
involving a quark loop. Right: NLO terms involving a gluon loop.}
\end{figure}

The Balitsky-Kovchegov (BK) equation  \cite{Balitsky:1995ub,Kovchegov:1999yj}
describes the pQCD evolution with increasing energy
of the forward scattering amplitude for the scattering between a quark-antiquark dipole
and a generic hadronic target (another dipole, a proton, or a nucleus), in the limit where
the number of colors is large ($N_c\to\infty$). The leading order (LO) version of
this equation defines the `leading logarithmic approximation' (LLA): it
resums all radiative corrections in which each
power of the QCD coupling $\abar\equiv\alpha_s N_c/\pi$, assumed to be
fixed and small, is accompanied by the energy
logarithm $Y\equiv \ln(s/Q_0^2)$ (the `rapidity'), with $s$ the
center-of-mass energy squared and $Q_0$ the characteristic transverse
scale of the target. The LO BK equation reads
\begin{align}\label{BK}
 \frac{\del T_{\bx\by}(Y) }{\del Y}=
 \frac{\abar}{2\pi}\, \int \rmd^2\bz\,
\frac{(\bx \minus \by)^2}{(\bx \minus \bz)^2 
 	(\bz \minus \by)^2} 
\, \big[ \minus T_{\bx\by}(Y) +T_{\bx\bz}(Y)+ T_{\bz\by}(Y)  \minus
 T_{\bx\bz}(Y) T_{\bz\by}(Y)
 \big]\,.
\end{align}
where the subscripts $\bx,\,\by,\,\bz$ denote the transverse coordinates 
of the original quark-antiquark pair and,
respectively, the soft gluon emitted in one step of the evolution. 
The first term in the r.h.s., which is negative, is a `virtual' correction where the
soft gluon has no overlap with the target. The other 3 terms describe `real' corrections
where the virtual gluon exists at the time of scattering (see the left diagrams
in Fig.~\ref{fig:diag}). In particular, the term quadratic
in $T$, which is negative, describes unitarity corrections associated with multiple scattering;
these become important when the target looks dense on the resolution scales of the
projectile. For what follows though, we shall be mostly interested in the dilute target,
or weak scattering, regime, where this quadratic term is negligible and \eqn{BK}
can be well approximated by its linearized version, the celebrated BFKL equation.
We shall moreover focus on the situation where the dipole looks very small
on the transverse scale of the target: $rQ_0\ll 1$, or $Q^2\gg Q_0^2$, with $r\equiv
|\bx-\by| \equiv 1/Q$. Indeed, this regime is characterized by
the existence of large radiative corrections,
enhanced by the {\em transverse} (or `collinear') logarithm $\rho\equiv\ln(Q^2/Q_0^2)$.
These corrections come from gluons emissions
which occur far outside the original dipole, such that $r\ll |\bx-\bz|\simeq |\bz-\by|\ll 1/Q_0$.
Such gluons look soft compared to their parent dipole but still hard compared to
the target, so they scatter only weakly: $T(z)\ll 1$.
In this regime,  $T(z)\sim z^2$, hence 
the (linear) `real' terms in \eqn{BK} dominate over the `virtual' one:
\beq\label{DLA}
  \frac{\del }{\del Y}\,\frac{T(r,Y)}{r^2}
 	\simeq \abar \int_{r^2}^{1/Q_0^2}
 	\dif{z}^2\, \frac{r^2}{{z}^2}\,\frac{T(z,Y)}{z^2}
\,.\eeq
The solution to this equation resums powers of $\abar Y\rho$
to all orders. This double logarithmic enhancement --- an energy logarithm
and a collinear one --- reflects the soft and collinear
singularities of bremsstrahlung. But \eqn{DLA} is not yet the correct
double-logarithmic approximation in QCD at high energy, as we shall see.

The next-to-leading order 
(NLO) corrections to  \eqn{BK} arise from 2-loop diagrams
which involve at least one soft gluon (see Fig.~\ref{fig:diag}).
The maximal contribution {\em a priori} expected for such a diagram 
(after subtracting the respective LLA piece, if any) is of order
$(\abar Y\rho)\times(\abar \rho)=\abar^2 Y\rho^2$; such a contribution
would provide a NLO correction $\sim\abar\rho$ to the BFKL kernel which is 
enhanced by a collinear log.
Yet, the explicit calculation of all such 2-loop graphs in Ref.~\cite{Balitsky:2008zza} 
reveals the existence of even larger corrections, of relative order $\abar\rho^2$,
which are enhanced by a {\em double} collinear logarithm. The complete result at
NLO appears to be extremely complicated  \cite{Balitsky:2008zza},
but it drastically simplifies if one keeps only the terms which are enhanced
by at least one transverse logarithm in the regime where $Q^2\gg Q_0^2$. Then 
it reads (at large $N_c$) 
 \beq\label{NLO}
\frac{\del T(r,Y)}{\del Y}
 	\simeq \abar \int_{r^2}^{1/Q_0^2}
 	\dif{z}^2\, \frac{r^2}{{z}^4}
 	\left\{1 
 	-\abar\left(\frac{1}{2}\ln^2 \frac{{z}^2}{r^2} +\frac{11}{12}
 	\ln \frac{{z}^2}{r^2} -\bar{b}\, \ln r^2 \mu^2  \right)\right\} T({z},Y)\,,
\eeq
which exhibits 3 types of NLO terms: the double-collinear log previously mentioned, 
a single collinear log which can be recognized as part of the DGLAP evolution (see below),
and the one-loop running coupling. ($\bar{b} = (11\Nc - 2 \Nf)/12\Nc$ is the first coefficient of the 
QCD $\beta$-function, and $\mu$ is a renormalization scale at which the coupling is evaluated.)
The NLO corrections enhanced by collinear logs are negative and large and lead to
numerical instabilities which render the NLO BK equation void of any predictive power
\cite{Lappi:2015fma,Iancu:2015vea}. The main source of this difficulty is the double-collinear
logarithm (DCL) $\abar\rho^2$, whose origin and resummation will be discussed in the next sections.

\section{Time ordering and double-collinear logarithms}
\label{sec:to}

The NLO correction $\sim\abar\rho^2$ to the kernel arises from a particular
2-loop contribution of order $\abar^2 Y\rho^3$, which looks anomalously large: it
involves a total of 4 (energy or transverse) logarithms, like the respective
LLA contribution $\sim(\abar Y\rho)^2$. As a matter of facts, this particular NLO
contribution is generated by the same 2-loop diagrams (in terms of topology and
kinematics) that are responsible for 2 successive steps in the LLA evolution described by
\eqn{DLA}:
namely, Feynman graphs involving 2 gluon emissions which are strongly
ordered in both longitudinal momentum and transverse momentum
(or transverse size). The physical interpretation
of the enhanced contributions becomes most transparent when the 
2-loop diagrams are computed within light-cone, or time-ordered,
perturbation theory \cite{Iancu:2015vea}.

\begin{figure*}
\centerline{\begin{minipage}[b]{0.35\textwidth}
\begin{center}
\includegraphics[width=0.95\textwidth,angle=0]{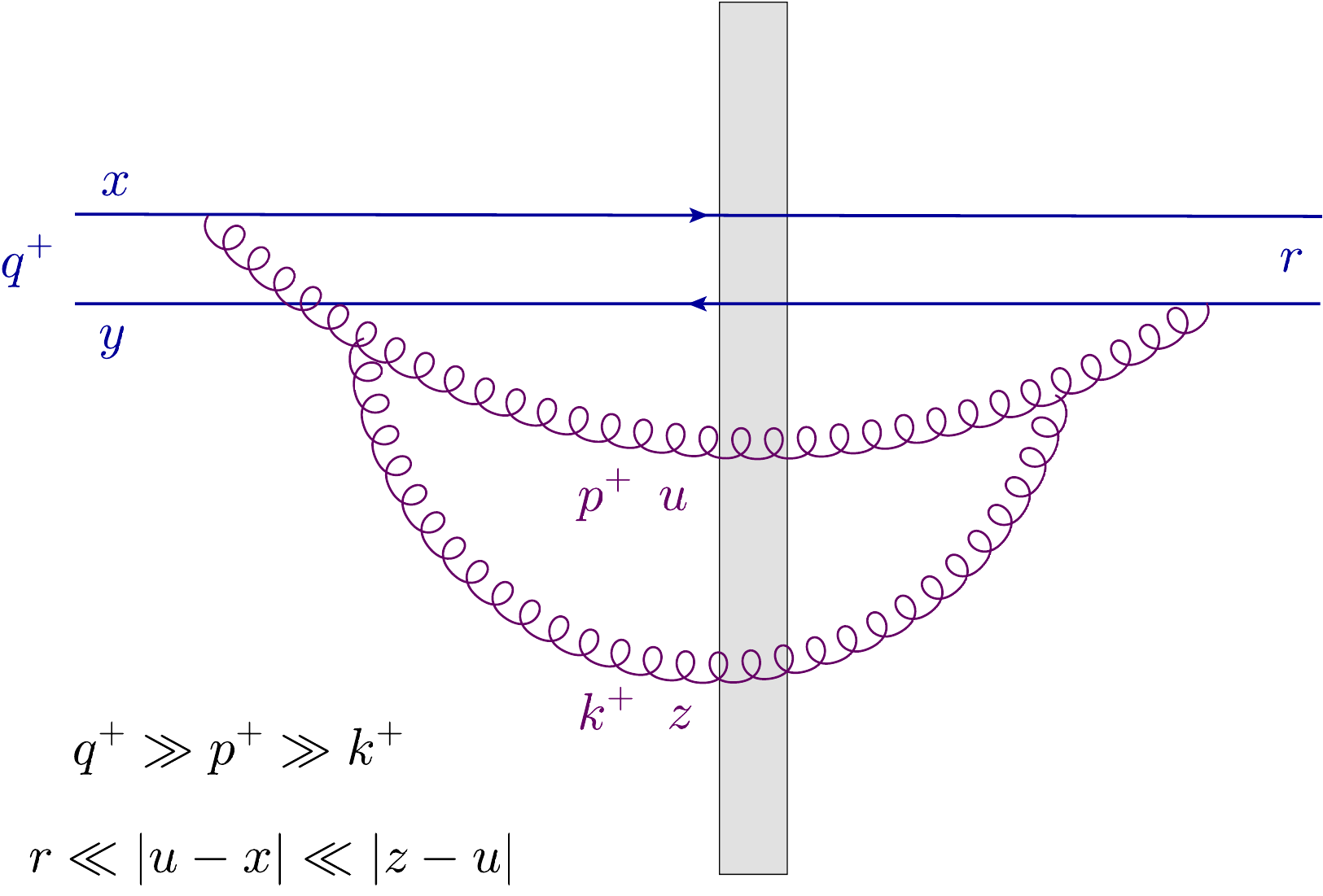}
\end{center}
\end{minipage}
\begin{minipage}[b]{0.3\textwidth}
\begin{center}
\includegraphics[width=0.9\textwidth,angle=0]{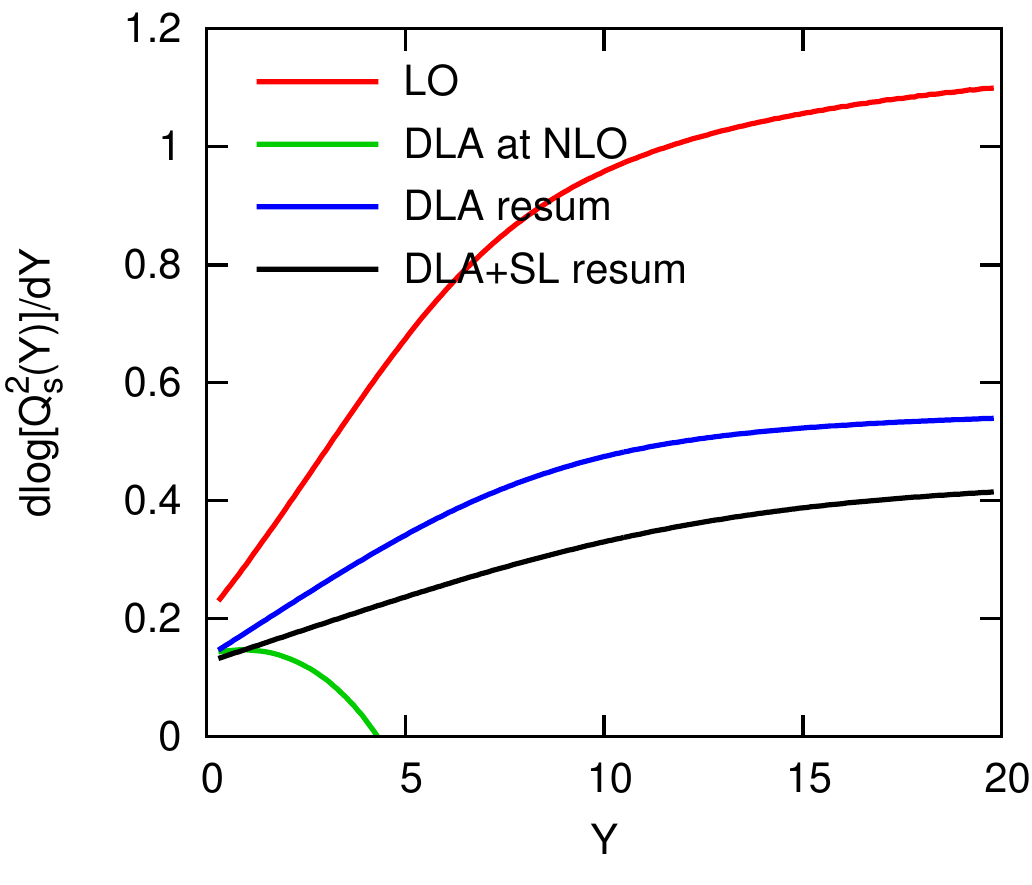}
\end{center}
\end{minipage}
\begin{minipage}[b]{0.33\textwidth}
\begin{center}
\includegraphics[width=0.9\textwidth,angle=0]{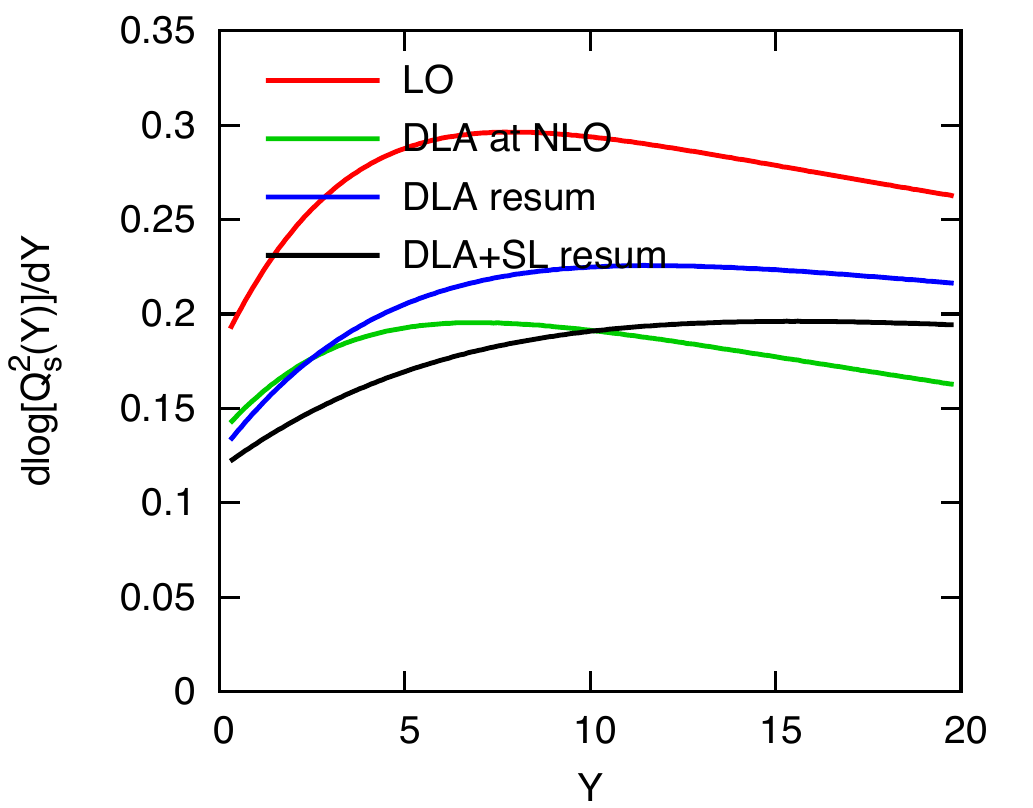}
\end{center}
\end{minipage}
}
\caption{\label{fig:nlodiag} Left: a typical diagram yielding a NLO correction enhanced by a DCL.
Middle and right: the saturation exponent $\lambda_s \equiv \dif \ln Q_s^2 /\dif Y$ as predicted by the
resummed equation \eqref{RGBK} with fixed  coupling $\abar=0.25$ 
(middle) and respectively running  coupling (right). The `DLA at NLO' curves show the
instability of the strict NLO approximation, whereas the curves denoted as
`DLA resum' and `DLA+SL resum'
demonstrate the effect of successive resummations in stabilizing and slowing down the evolution.}
\end{figure*}

For example, let us examine the diagram in Fig.~\ref{fig:nlodiag} (left) where the 
longitudinal momenta of the emitted gluons obey $q^+\gg p^+\gg k^+$,
whereas their transverse sizes are ordered according to
\beq\label{strord}
r=|\bx-\by|\,\ll\,|\bu-\bx|\simeq |\bu-\by|\,\ll\,|\bz-\bx|\simeq
 |\bz-\by|\simeq |\bz-\bu|\,\ll\,1/Q_0\,.\eeq
We implicitly assumed here that the dipole projectile is a right
mover with large longitudinal momentum $q^+$, while 
the hadronic target is a left mover with large momentum $P^-$.
As visible in Fig.~\ref{fig:nlodiag} (left), the softer gluon $k$ is emitted after and absorbed 
before the harder one $p$. This particular time-ordering introduces 
the energy denominator $1/(k^-+p^-)$ which in turn implies that 
the largest logarithmic contributions occur when the {\em lifetimes}
of the two gluons are also strongly ordered: $\tau_k \equiv 2k^+/\bk^2 \ll \tau_p \equiv 2p^+/\bp^2$. 
Here $\bp$ is the transverse momentum of the gluon $p$, related to its transverse size
via the uncertainty principle, $\bp^2\sim 1/(\bu-\bx)^2$, and similarly for the gluon $k$.
Indeed, when this condition $\tau_k \ll  \tau_p$ is satisfied, then the four integrations
over $p^+$, $u$, $k^+$, and $z$ are all logarithmic\footnote{Other diagrams which are not time-ordered may contain double logarithms individually, but they cancel in the final answer \cite{Iancu:2015vea}.}. Different hookings of the two gluons lead to 32 diagrams like the one in Fig.~\ref{fig:nlodiag} (left). 
Adding all of these contributions, we 
find in the regime defined in \eqn{strord} (with simplified notations $|\bu-\bx|\to u$
and $|\bz-\bu|\to z$)
 \begin{equation}
 \label{deltat12}
 	\Delta T(r) = \abar^2\int\frac{\dif k^+}{k^+}\frac{\dif{z}^2}{z^2}
 	\int \frac{\dif p^+}{p^+}\frac{\dif{u}^2}{u^2} 	\Theta \left(p^+ u^2 -k^+ z^2 \right)
	\, \frac{r^2}{z^2}\,T(z)\,,
 \end{equation}
where the step-function implements the lifetime constraint. If there were not for this constraint,
\eqn{deltat12} would look identical as two iterations of the LO equation \eqref{DLA}.
By integrating out the intermediate gluon $p^+$, 
\beq
\abar\int_{r^2}^{z^2}\!\frac{\dif{u}^2}{u^2} \!\int_{k^+}^{q^+} \!\frac{\dif p^+}{p^+}\,
 \Theta \left(p^+ u^2 -k^+ z^2 \right)\,=\,
\abar\left(\ln\frac{z^2}{u^2}\ln\frac{q^+}{k^+}-\frac{1}{2}\ln^2\frac{z^2}{u^2}\right)\,=\,
\abar Y\rho-\frac{\abar\rho^2}{2}\,,
\eeq
one finds the expected LLA contribution $\abar Y\rho$ 
plus a term independent of $Y$, namely $-{\abar\rho^2}/{2}$, to be interpreted as a NLO
correction to the kernel for emitting the softer gluon $k^+$. This correction 
reproduces the DCL part of the NLO correction in \eqn{NLO}, thus clarifying the
physical interpretation of the latter: it expresses the reduction in the 
rapidity interval $\Delta Y$ available to the intermediate gluon due to the time-ordering constraint.
This argument extends to all orders  \cite{Iancu:2015vea}:  the perturbative corrections enhanced
by DCLs can be resummed to all orders by enforcing time-ordering within the
 `naive' LLA.  However this procedure has the drawback to produce an evolution
equation which is {\em non-local} in rapidity \cite{Iancu:2015vea,Beuf:2014uia},
as already manifest in \eqn{deltat12}. This non-locality reflects the fact that the
natural evolution variable at high energy is not the longitudinal momentum
$k^+$ of a gluon from the projectile, or the associated rapidity $Y=\ln(q^+/k^+)$, 
but rather its lifetime $\tau_k = 2k^+/k_\perp^2$,
or equivalently $\eta\equiv Y-\rho$ with $\rho=\ln(Q^2/k_\perp^2)$:
the evolution is local in $\eta$, but not in $Y$.

\section{The collinearly-improved BK equation}
\label{sec:RGBK}

At this stage, something remarkable happens: the non-local equation with time-ordering
can be {\em equivalently} rewritten as an equation {\em local} in $Y$, where however
both the kernel and the initial condition at $Y=0$ resum corrections to all orders in $\abar\rho^2$
\cite{Iancu:2015vea}. This equation can furthermore be extended to resum the {\em single} transverse
logarithms that appear at NLO, cf. \eqn{NLO}, namely the single collinear logarithms (SCL) 
which represent the beginning of the  DGLAP evolution
and the one-loop running coupling corrections \cite{Iancu:2015joa}.

The SCL too arises from successive emissions in which the second gluon is much softer, both in transverse and longitudinal momenta, than the first one, but now it is the region $\tau_k \sim \tau_p$ which gives the relevant contribution. Its coefficient $A_1=11/12$ can be recognized as the first non-singular term in the small-$\omega$ expansion of the DGLAP anomalous dimension  \cite{Iancu:2015joa}.   
This implies that in order to resum such SCLs, it suffices to include $A_1$ as an `anomalous dimension', i.e.~as a power-law suppression in the evolution kernel. The running coupling corrections can be
resummed by choosing the renormalization scale $\mu$ as the hardest
scale in the problem: 
$\abar\to \abar(r_{\rm min})$, where $r_{\rm min}$ is the size of the smallest dipole,
$r_{\rm min} \equiv \min\{|\bx \minus\by|,|\bx \minus\bz|,|\by \minus\bz|\}$.

We are thus led to the following, {\em collinearly-improved}, version of the BK equation, which 
faithfully includes the NLO effects enhanced by large (double or single) transverse logarithms,
but improves over the strict NLO approximation by resumming similar corrections to all orders:
\begin{align}\label{RGBK}
 	\frac{\del T_{\bx\by}}{\del Y} = 
 	\int  \frac{\dif^2 \bz}{2\pi}\,  \abar(r_{\rm min})\,
 	\frac{(\bx \minus \by)^2}{(\bx \minus \bz)^2 
 	(\bz \minus \by)^2}\,&
	\left[\frac{r^2}{z^2_<}\right]^{\pm A_1 \abar } 
	\mathcal{K}_{\sdla}\big(\bar \rho^2\big)\, 
	\big[ \minus T_{\bx\by} +T_{\bx\bz}+ T_{\bz\by}  \minus T_{\bx\bz} T_{\bz\by} \big].
 \end{align}
In this equation, $z^2_< \equiv \min\{(\bx \minus \bz)^2,(\by \minus \bz)^2\}$, \,
$\bar \rho^2\equiv \ln[(\bx \minus\bz)^2/r^2] \ln [(\bz \minus\by)^2/r^2]$, and 
(with $\rmJ_1$ the Bessel function) 
 \begin{equation}\label{kdla}
 	\mathcal{K}_{\sdla}(\rho^2) \equiv \frac{\rmJ_1
 	\big(2\sqrt{\abar \rho^2}\big)}{\sqrt{\abar \rho^2}} = 
 	1- \frac{\abar \rho^2}{2} + \frac{(\abar\rho^2)^2}{12} + \cdots
 \end{equation}
is the change in the kernel accounting for DCLs to all orders.
A similar kernel was found in transverse momentum space \cite{Vera:2005jt}, as an 
approximation to resummations performed in the context of the BFKL equation.
The initial condition to \eqn{RGBK} at $Y=0$ can be found in Refs.~\cite{Iancu:2015vea,Iancu:2015joa},
to which we refer for more details.

In contrast to the NLO BK equation, the resummed equation \eqref{RGBK} admits stable 
solutions, which are well-suited for phenomenology. 
The various resummations considerably slow down the evolution, so the ensuing  ``evolution speed'' --- as measured by the saturation exponent $\lambda_s \equiv \dif \ln Q_s^2 /\dif Y$, with $Q_s(Y)$
the saturation momentum
--- is substantially smaller than that predicted by the LO evolution (see Fig.~\ref{fig:nlodiag}). 
By using  \eqn{RGBK} together with appropriate forms for the initial condition, we have been able to
obtain good quality fits to the HERA data \cite{Aaron:2009aa}
for the $ep$ reduced cross section at small Bjorken $x \leq 10^{-2}$, with only 4 free parameters
 (for similar fits, without inclusion of the SCLs, see \cite{Albacete:2015xza}).
The evolution speed extracted from the fits is $\lambda_s = 0.20 \div 0.24$. 
A remarkable feature about these fits is that they
are rather discriminatory: they exclude several models for the initial conditions
previously used in the literature and also
some previous choices for the running coupling. Conversely,
they favor the running-coupling version of the McLerran-Venugopalan model 
for the initial condition and the `smallest dipole' prescription $\bar{\alpha}(r_{\rm min})$
for the running of the coupling.

This work is supported by the European Research Council under the Advanced Investigator Grant ERC-AD-267258 and by the Agence Nationale de la Recherche project \# 11-BS04-015-01. The work of A.H.M.
is supported in part by the U.S. Department of Energy Grant \# DE-FG02-92ER40699.











\providecommand{\href}[2]{#2}\begingroup\raggedright\endgroup

\end{document}